\documentclass[12pt]{article}
\usepackage{epsfig}
\def\be{\begin{equation}}
\def\ee{\end{equation}}
\def\bea{\begin{eqnarray}}
\def\eea{\end{eqnarray}}
\usepackage{graphicx}% Include figure files
\usepackage{caption}
\usepackage{hyperref}
\usepackage{float}
\usepackage{amsmath}

\catcode`\@=11
\def\lsim{\mathrel{\mathpalette\@versim<}}
\def\gsim{\mathrel{\mathpalette\@versim>}}
\def\@versim#1#2{\vcenter{\offinterlineskip
\ialign{$\m@th#1\hfil##\hfil$\crcr#2\crcr\sim\crcr } }}
\catcode`\@=12
\usepackage{axodraw}

\catcode`@=12
\begin{document}
\thispagestyle{empty}
\begin{flushright}
UCRHEP-T545\\
October 2014\
\end{flushright}
\vspace{0.6in}
\begin{center}
{\LARGE \bf Dark Matter with Flavor Symmetry\\ 
and its Collider Signature\\}
\vspace{1.2in}
{\bf Ernest Ma and Alexander Natale\\}
\vspace{0.2in}

{\sl Department of Physics and Astronomy, University of California,\\
Riverside, California 92521, USA\\}
\end{center}
\vspace{1.2in}
\begin{abstract}\
The notion that dark matter and standard-model matter are connected through 
flavor implies a generic collider signature of the type 2 jets + $\mu^\pm$ 
+ $e^\mp$ + missing energy.  We discuss the theoretical basis of this 
proposal and its verifiability at the Large Hadron Collider.
\end{abstract}

\newpage
\baselineskip 24pt

A generic framework\cite{m14} has been proposed for understanding how 
dark matter (DM) and flavor are connected through the 125 GeV 
particle\cite{atlas12,cms12} discovered at the Large Hadron Collider (LHC).
It is assumed to be the one Higgs boson $h$ of the standard model (SM), but 
its couplings to some of the SM fermions are forbidden at tree level by a 
flavor symmetry such as $A_4$, and occur only in one loop by the soft 
breaking of this flavor symmetry in the dark sector.  A verifiable 
consequence is the possible deviation\cite{fm14} of the Higgs Yukawa 
coupling from the SM prediction of $m_f/v$, where $m_f$ is the mass of the 
fermion and $v=246$ GeV is the vacuum expectation value of $h$.  Here we 
consider a generic collider signature from the new particles of this proposal.

Our specific starting point is the radiative generation of charged-lepton 
and $d$ quark masses as proposed in Ref.\cite{m14}.
\begin{figure}[htb]
\vspace*{-3cm}
\hspace*{-3cm}
\includegraphics[scale=1.0]{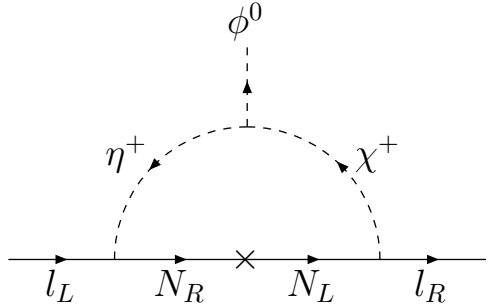}
\vspace*{-21.5cm}
\caption{One-loop generation of charged-lepton mass.}
\end{figure}
\begin{figure}[htb]
\vspace*{-3cm}
\hspace*{-3cm}
\includegraphics[scale=1.0]{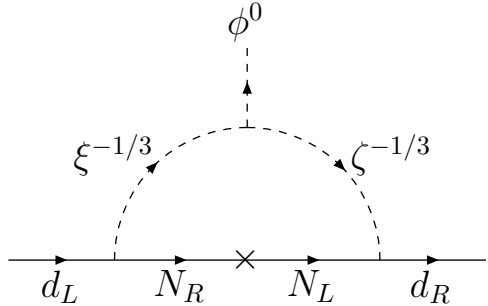}
\vspace*{-21.5cm}
\caption{One-loop generation of $d$ quark mass.}
\end{figure}

In the above, $\eta^+$ is part of a scalar electroweak doublet 
$(\eta^+,\eta^0)$ first introduced in Ref.\cite{m06}.  It is distinguished 
from the SM Higgs doublet $(\phi^+,\phi^0)$ by an exactly conserved (dark) 
discrete $Z_2$ symmetry, under which $(\eta^+,\eta^0)$ is odd and 
$(\phi^+,\phi^0)$ is even.  The scalar singlet $\chi^+$ is also odd. 
There are three neutral singlet Dirac fermions $N_{1,2,3}$ which are odd 
under $Z_2$ as well.  The scalar color triplet $\xi^{-1/3}$ is part of an 
electroweak doublet $(\xi^{2/3},\xi^{-1/3})$ and $\zeta^{-1/3}$ is an 
electroweak singlet.  They are also odd under $Z_2$.  Hence all 
the particles in the loop are distinguished from those of the SM by this dark 
$Z_2$ symmetry.  If desired, $Z_2$ may be promoted to a gauged $U(1)_D$ 
as shown in Ref.\cite{mpr13}.  In this scheme, flavor is being 
carried by the neutral singlet fermions $N_{1,2,3}$, the lightest of which, 
say $N_1$, is a DM candidate.  

The usual tree-level SM Yukawa couplings, i.e. $\phi^0 \bar{l}_L l_R$ and 
$\phi^0 \bar{d}_L d_R$, are assumed to be forbidden by a flavor 
symmetry\cite{m14}.  This flavor symmetry is then broken softly by the 
$3 \times 3$ $\bar{N}_L N_R$ mass matrix, allowing the loop to be connected. 
Flavor is thus carried by the dark matter fermions $N_{1,2,3}$.  The 
flavor structure of dark matter is transmitted to the visible sector 
through the radiative mass-generating mechanism, using the one Higgs 
doublet of the standard model. 
The color-triplet scalars $\xi$ and $\zeta$ are analogs of the scalar 
quarks of supersymmetry (SUSY), but there is only one copy of each 
and they do not carry flavor.  Because they are colored, they are 
produced copiously in pairs 
by gluons at the Large Hadron Collider (LHC).  To see how they may be 
detected, consider the following scenario with the simplifying assumed 
interactions:
\begin{equation}
{\cal L}_{int} = f (\bar{d}_R N_{1L} + \bar{s}_R N_{2L}) \zeta^{-1/3} 
+ f' (\bar{e}_R N_{1L} + \bar{\mu}_R N_{2L}) \chi^- + H.c.
\end{equation}
Assume further that $m_\zeta > m_{N_2} > m_\chi > m_{N_1}$, then $\zeta$ decays to 
$s N_2$ and $d N_1$.  Whereas $N_1$ is stable, $N_2$ decays to $\mu^\pm 
\chi^\mp$, and $\chi^\mp$ decays to $e^\mp N_1$.  This implies a signature 
of the type 2 jets + $\mu^\pm$ + $e^\mp$ + missing energy ($E_T^{\rm{miss}}$) 
at the LHC, and is rather distinct 
because of the different charged leptons in the final state.  In 
supersymmetry, flavor is organized from quark to squark and lepton 
to slepton.  Here it is organized through the flavored DM 
particles $N_{1,2,3}$.

%Methodology:

In the following we take the above simplified model, and see how it may be 
probed at the LHC.  The color-triplet $\zeta$ behaves as a squark, so it 
couples to gluons, but since there is no gluino, it has no connection 
to quarks except through Eq.~(1) which always involves $N_{1,2,3}$.  As such, 
the branching fractions of $\zeta \to d N_1$ and $\zeta \to s N_2$ are 
roughly equal, i.e. 0.5 each.  The mass of the charged scalar $\chi$ is 
constrained to be greater than 70 GeV from LEP data \cite{chargedscalar}. 
However, after analysing the model with a range of masses, we find that 
the best scenarios of optimizing the ratio of the cross sections of the 
signal divided by background under various cuts are those for $m_{N_2} = 
400$ GeV and $m_{\chi} = 200$ GeV.  For the results below, we use the 
specific mass scheme of $m_{\zeta} > m_{N_2} > m_{\chi} > m_{N_1}$, as already 
mentioned.

Our model is implemented in CalcHEP \cite{calchep} to generate parton-level 
events using the CTEQ6M parton distribution functions (PDF) \cite{PDF}, which 
are then analysed with PYTHIA 8 \cite{pythia6, pythia8} to produce 
leading-order (LO) results.  The LO production cross section of the 
squark analogs (hereby referred to simply as squarks) is verified through 
the Feynrules \cite{feynrules} interface with Madgraph 5 \cite{madgraph5nlo}, 
producing a cross section consistent with CalcHEP.  Whereas the main signature 
of this model is distinct from that of SUSY squarks, SUSY models with only 
one light family of squark and the gluinos decoupled (called simplified 
topologies) will have the same production cross section as the squarks here. 
Most importantly, the masses excluded by the LHC are much lower for such 
models as seen in Fig. 3.  This scenario is used for the expected 13 TeV 
data where the production cross sections of the squarks are compared to 
simplified topology models of SUSY squarks, which are calculated at 
Next-To-Leading-Order (NLO) and Next-Leading-Log (NLL) by 
C. Borschensky \textit{et al.} \cite{NLOSquark}.  The comparison of our 
LO calculation to these results is used to obtain a k-factor in order 
to approximate the NLO contributions to the squark production.  For the 
opposite-sign opposite-flavor dilepton events, the main background is 
from $t \bar{t}$ pairs, unlike the same-flavor case which has significant 
contribution from Drell-Yan production\cite{dilepton8tev}.  For the 
expected 13 TeV data, only the $t \bar{t}$ background is generated 
with CalcHEP, using a k-factor to scale to the NLO production cross 
section for $t \bar{t}$  \cite{ttbar}, and analysed with PYTHIA 8.

In addition to the opposite-sign opposite-flavor dilepton + 2 jets + 
$E_T^{\rm{miss}}$ signature, it is also possible for each squark to decay 
directly to DM and a quark, thus producing two jets and missing energy, without 
any lepton.  As a result, SUSY searches at 7 TeV and 8 TeV for this 
signature in simplified SUSY topologies offer useful constraints on our model. 
The searches at 8 TeV \cite{squarkLSP8tev} are presented in Fig. 3.  For 
our model, the 7 TeV (not shown) and 8 TeV (Fig. 3) data are taken into 
account by ensuring the cross section for $\zeta$ decaying directly to 
$d N_1$ is lower than the upper limit observed at the LHC for a single 
squark (in a simplified topology) decaying directly to a quark + LSP. After 
these constraints are taken into account, the results from the 7 TeV 
\cite{dilepton7tev} and 8 TeV \cite{dilepton8tev} searches looking for 
events with 2 leptons, 2 jets, and missing energy do not provide any 
further constraints.  Additionally, it is possible for the squark to be produced through a t-channel process directly with DM producing a monojet signal, however this production cross section is a smaller contributor to an LHC signal than the dijet + $E_T^{\rm{miss}}$.  Such a monojet + $E_T^{\rm{miss}}$ signature for DM has been investigated in a model independent way (see \cite{SqrkDD2,8TeVMJ,SqrkDD,fpRelic}).  When taking into account the monojet signature, the upper bound of allowed events in the 8 TeV data \cite{8TeVMJ} are taken into account, at LO, if a squark mass above 400 GeV is assumed. Additionally, studies on DM that can interact with a colored scalar have explored the constraints from relic abundance and direction detection of DM, which are potentially more restrictive than the LHC \cite{SqrkDD2,SqrkDD,fpRelic}.  In particular, XENON100 is able to probe down to $10^{-45}$ $cm^2$ for a DM mass of 100 GeV \cite{fpRelic}, which would rule out much of the parameter space if $f$ is of order unity.  However, to yield the proper down quark mass a value of $f \approx 0.01 $ must be used and the spin dependent, direct detection, cross section for Dirac fermion dark matter \cite{fpRelic} can be of order $10^{-45}$ $cm^2$ for a squark mass of 400 GeV and a DM mass of 100 GeV.

%Setup at 13 TeV:

We now present our analysis for the expected 13 TeV run.  Six cuts are 
applied to the signal and background events in PYTHIA, with four of the 
cuts corresponding to the cut regions from \cite{dilepton7tev}, while the 
last two cut regions are found to be effective for our model based on our 
analysis.  All of the six cuts are described in TABLE 1 below, with the 
resulting $t \bar{t}$ decay cross section in each cut region.  Each cut 
is implemented in PYTHIA 8 and applied to both the signal events, and the 
background events from $t \bar{t}$ decays.  A signal-to-background (SB) 
ratio of the resulting cross sections is calculated for each choice of 
squark and DM mass.  In Figs. 4 and 5 we show the regions in which the choice 
of DM mass and squark mass satisfies $SB > 5$ for various cuts.  Two of the 
cut regions, R1 and R4, do not have any mass choice for which $SB > 5$, 
and so do not appear in Figs. 4 and 5.

%\begin{itemize}
%\item{} R1: $E_T^{\text{miss}} >$ 275 GeV, $H_T >$ 300 GeV,
%\item{} R2: $E_T^{\text{miss}} >$ 200 GeV, $H_T >$ 600 GeV,
%\item{} R3: $E_T^{\text{miss}} >$ 275 GeV, $H_T >$ 600 GeV,
%\item{} R4: $E_T^{\text{miss}} >$ 275 GeV, 125 $< H_T < $ 300 GeV,
%\item{} R5: $E_T^{\text{miss}} >$ 200 GeV, $H_T >$ 350 GeV,
%\item{} R6: $p_T^{\text{jet}} >$ 150 GeV, $p_T^{\text{lep}} >$ 25 GeV, $E_T^{\text{miss}} >$ 200 GeV, $H_T >$ 350 GeV.
%\end{itemize}

%Results:

\begin{table}[H]
\caption{Cuts applied to the signal and background for opposite-sign 
opposite-flavor dileptons + 2 jets + missing energy ($E_T^{\text{miss}}$). The 
values of $E_T^{\text{miss}}$, $H_T$ (the scalar sum of transverse jet momentum), 
and transverse momentum of jets and leptons ($p_T$) are in GeV.  Also shown 
are the resulting SM background cross section after the cuts are 
applied in PYTHIA.}
\begin{tabular}{c c c c c c c}
\hline \hline
Cut: & $E_T^{\text{miss}}$ & $H_T$ & $p_T^{\text{j}}$ ($p_T^{\text{l}}$) & 
$|\eta_{\text{j}}|$ upper-limit & $|\eta|$ e ($\mu$) upper-limit & 
$\sigma_{\text{post-cut}}$ (fb)\\
\hline
R1 & 275 & 300 & 30 (20) & 3.00 & 2.40 (2.50) & 10.0 \\
R2 & 200 & 600 & 30 (20) & 3.00 & 2.40 (2.50) & 0.5 \\
R3 & 275 & 600 & 30 (20) & 3.00 & 2.40 (2.50) & 0.4 \\
R4 & 200 &  $>$ 125, $<$ 300 & 30 (20)  & 3.00 & 2.40 (2.50) & 33.1 \\
R5 & 200 & 350 & 30 (20) & 3.00 & 2.40 (2.50) & 7.1 \\
R6 & 200 & 350 & 150 (25) & 3.00 & 2.40 (2.50) & 1.2 \\
\hline \hline
\end{tabular}
\end{table}
\label{table:cuts}

As seen in Figs. 4 and 5, the cuts R5 and R2 allow fewer mass choices to have 
a large SB ratio.  This can be understood after consulting the resulting 
background cross sections in TABLE 1, which show that the background cross 
section for these cuts are larger than the cuts R6 and R3, so while fewer 
background events survive these more stringent cuts, the background events 
are cut down even further producing the results seen in the figures.

In conclusion, the model outlined in this paper could be observed at the LHC 
during the 13 TeV run, and has a signature distinct from SUSY. The major 
difference between this model and SUSY is that the signature is produced 
solely in the opposite-flavor channel, however, same-sign searches use the 
opposite-flavor events to estimate the flavor symmetric background 
\cite{dilepton8tev}, and subtract it from the observed same-sign background 
to obtain a signal for SUSY \cite{dilepton8tev}. Given a similar search 
strategy, our model would predict a significant negative signal in 
same-flavor searches.  As a result, any large, positive, signal in the 
same-flavor channel could potentially rule out or heavily constrain our model. 
For example the mass choices that produce large SB ratios would be ruled out 
in such a scenario.  In addition, searches at 13 TeV for $\zeta$ decaying 
directly to $d N_1$ will provide further constraints.

\underline{Acknowledgement} :  This work is supported in part 
by the U.~S.~Department of Energy under Grant No.~DE-SC0008541.

\baselineskip 18pt

\bibliographystyle{unsrt}

\begin{figure}[htb]
\begin{center}
\includegraphics[scale=0.8]{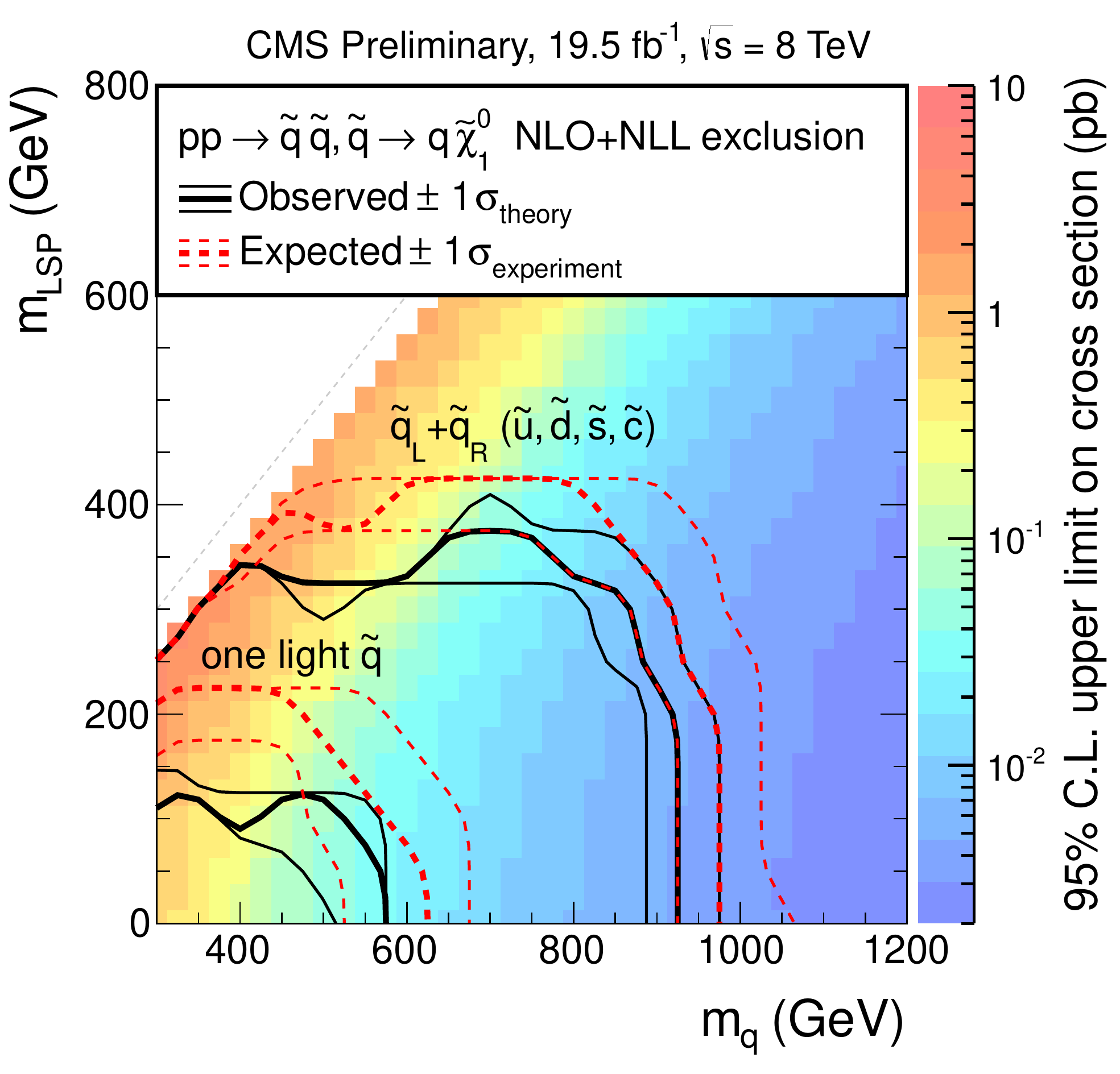}
\caption{Results of squark searches at the LHC of squark decay to quark + 
LSP at 8 TeV  from CMS-PAS-SUS-13-019 \cite{squarkLSP8tev}.}
\end{center}
\end{figure}

\begin{figure}[htb]
\includegraphics[scale=1]{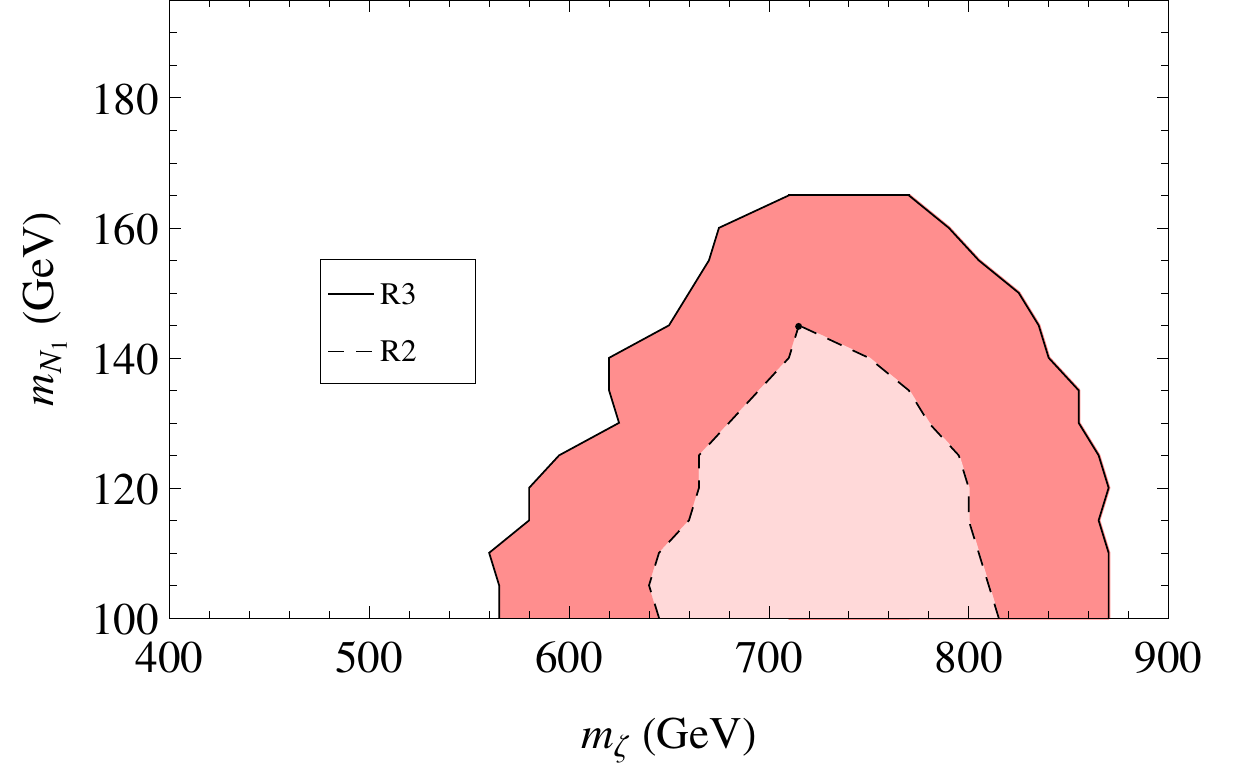}
\caption{Masses for $N_1$ and $\zeta$ that could produce a 
signal-to-background ratio, when compared to $ t \bar{t}$ decays, larger 
than 5 in the opposite-sign opposite-flavor dilepton + 2 jets + missing 
energy signature, under the R2 and R3 cuts.}
\end{figure}

\begin{figure}[htb]
\includegraphics[scale=1]{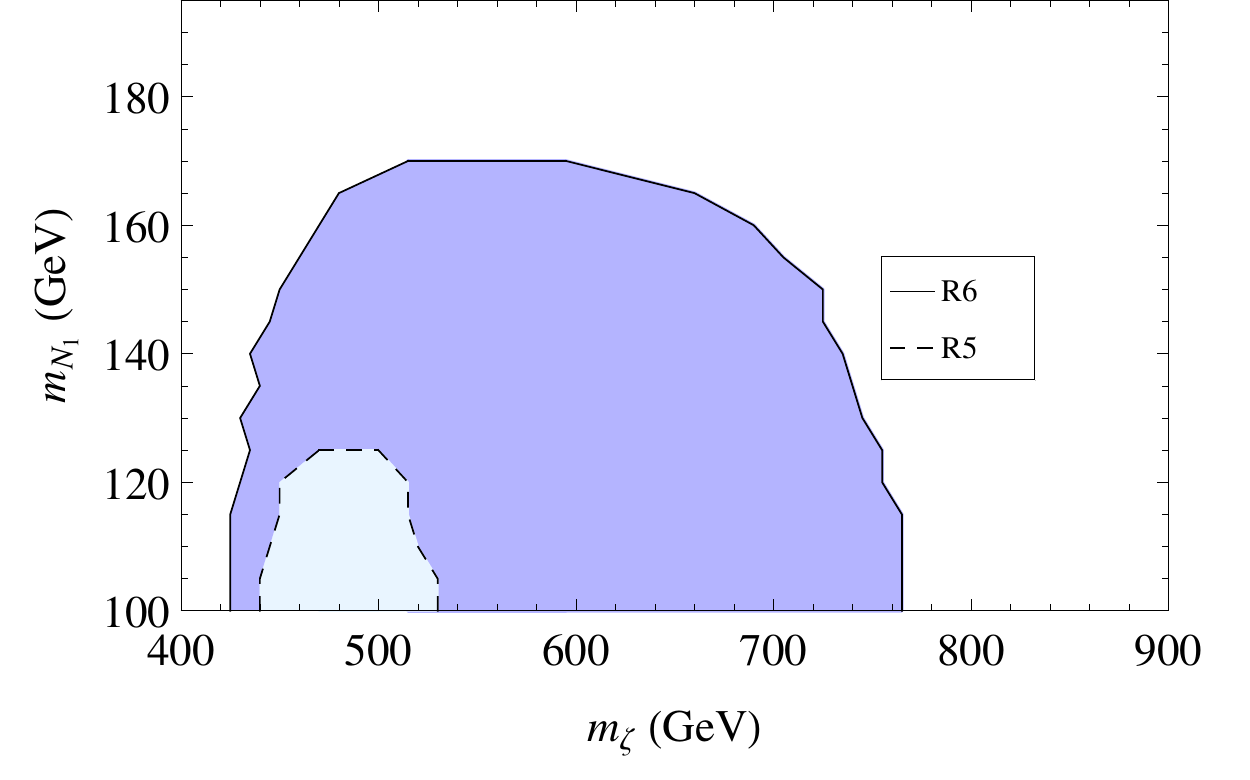}
\caption{Masses for $N_1$ and $\zeta$ that could produce a 
signal-to-background ratio, when compared to $ t \bar{t}$ decays, larger 
than 5 in the opposite-sign opposite-flavor dilepton + 2 jets + missing 
energy signature, under the R5 and R6 cuts.}
\end{figure}

\end{document}